\documentclass[superscriptaddress,aps,prx,twocolumn,longbibliography]{revtex4-1}
\usepackage{graphicx}
\usepackage{epstopdf}
\usepackage{dcolumn}
\usepackage{bm}
\usepackage{color}
\usepackage{amsmath}
\usepackage{ulem}
\usepackage{float}
\usepackage[colorlinks,linkcolor=blue, urlcolor=blue, anchorcolor=blue, citecolor=blue]{hyperref}

\usepackage{physics}

\begin{document}

\title{Absence of localized $5d^1$ electrons in KTaO$_3$ interface superconductors}
\author{Xinqiang Cai}
\affiliation{International Center for Quantum Materials, School of Physics, Peking University, Beijing 100871, China}

\author{Jungho Kim}
\affiliation{Advanced Photon Source, Argonne National Laboratory, Lemont, Illinois 60439, USA}

\author{Leonardo Martinelli}
\affiliation{Dipartimento di Fisica, Politecnico di Milano, Piazza Leonardo da Vinci 32, I-20133 Milano, Italy}

\author{Piero Florio}
\affiliation{Dipartimento di Fisica, Politecnico di Milano, Piazza Leonardo da Vinci 32, I-20133 Milano, Italy}

\author{Matteo Corti}
\affiliation{Dipartimento di Fisica, Politecnico di Milano, Piazza Leonardo da Vinci 32, I-20133 Milano, Italy}

\author{Weiliang Qiao}
\affiliation{International Center for Quantum Materials, School of Physics, Peking University, Beijing 100871, China}

\author{Yanqiu Sun}
\affiliation{School of Physics, Zhejiang University, Hangzhou 310027, China}

\author{Jiasen Niu}
\affiliation{Leiden Institute of Physics, Leiden University, 2333 CA Leiden, Netherlands}

\author{Quentin Faure}
\affiliation{ESRF—The European Synchrotron, 71 Avenue des Martyrs, CS 40220, F-38043 Grenoble, France}

\author{Christoph Sahle}
\affiliation{ESRF—The European Synchrotron, 71 Avenue des Martyrs, CS 40220, F-38043 Grenoble, France}

\author{Qingzheng Qiu}
\affiliation{International Center for Quantum Materials, School of Physics, Peking University, Beijing 100871, China}

\author{Qian Xiao}
\affiliation{International Center for Quantum Materials, School of Physics, Peking University, Beijing 100871, China}

\author{Xiquan Zheng}
\affiliation{International Center for Quantum Materials, School of Physics, Peking University, Beijing 100871, China}

\author{Qizhi Li}
\affiliation{International Center for Quantum Materials, School of Physics, Peking University, Beijing 100871, China}

\author{Changwei Zou}
\affiliation{International Center for Quantum Materials, School of Physics, Peking University, Beijing 100871, China}

\author{Xinyi Jiang}
\affiliation{International Center for Quantum Materials, School of Physics, Peking University, Beijing 100871, China}

\author{Giacomo Ghiringhelli}
\affiliation{Dipartimento di Fisica, Politecnico di Milano, Piazza Leonardo da Vinci 32, I-20133 Milano, Italy}

\author{Wei Han}
\affiliation{International Center for Quantum Materials, School of Physics, Peking University, Beijing 100871, China}
\affiliation{Collaborative Innovation Center of Quantum Matter, Beijing 100871, China}

\author{Yanwu Xie}
\affiliation{School of Physics, Zhejiang University, Hangzhou 310027, China}

\author{Yi Lu}
\affiliation{National Laboratory of Solid State Microstructures and Department of Physics, Nanjing University, Nanjing 210093, China}
\affiliation{Collaborative Innovation Center of Advanced Microstructures, Nanjing University, Nanjing 210093, China}

\author{Marco Moretti Sala}
\affiliation{Dipartimento di Fisica, Politecnico di Milano, Piazza Leonardo da Vinci 32, I-20133 Milano, Italy}

\author{Yingying Peng}
\email{yingying.peng@pku.edu.cn}
\affiliation{International Center for Quantum Materials, School of Physics, Peking University, Beijing 100871, China}
\affiliation{Collaborative Innovation Center of Quantum Matter, Beijing 100871, China}

\date{\today}

\begin{abstract}

Recently, an exciting discovery of orientation-dependent superconductivity was made in two-dimensional electron gas (2DEG) at the interfaces of LaAlO$_3$/KTaO$_3$ (LAO/KTO) or EuO/KTaO$_3$ (EuO/KTO). The superconducting transition temperature can reach a $T_c$ of up to $\sim$ 2.2 K, which is significantly higher than its 3$d$ counterpart LaAlO$_3$/SrTiO$_3$ (LAO/STO) with a $T_c$ of $\sim$ 0.2 K. However, the underlying origin remains to be understood. 
 To uncover the nature of electrons in KTO-based interfaces, we employ x-ray absorption spectroscopy (XAS) and resonant inelastic x-ray spectroscopy (RIXS) to study LAO/KTO and EuO/KTO with different orientations. We reveal the absence of $dd$ orbital excitations in all the measured samples. Our RIXS results are well reproduced by calculations that considered itinerant $5d$ electrons hybridized with O $2p$ electrons. This suggests that there is a lack of localized Ta $5d^1$ electrons in KTO interface superconductors, which is consistent with the absence of magnetic hysteresis observed in magneto-resistance (MR) measurements. These findings offer new insights into our understanding of superconductivity in Ta $5d$ interface superconductors and their potential applications.

\end{abstract}

\maketitle

\section{Introduction}

Oxide interfaces have attracted much attention since the discovery of conducting LaAlO$_3$/SrTiO$_3$ (LAO/STO) interfaces \cite{LAOSTO2DEG2004}. These interfaces give rise to intriguing emergent phenomena such as superconductivity \cite{LAOSTOSC2007}, ferromagnetism \cite{LAOSTOSCFM2011, LAOSTOFMSC2011} and electric-field tuned phase transitions \cite{LAOSTOEtuning2006, LAOSTOEtuningN2008, LAOSTOEtuningNP2008, LAOSTOEtuning2009}. Recently, the discovery of a superconducting two-dimensional electron gas (2DEG) in LaAlO$_3$(or EuO)/KTaO$_3$ (LAO/KTO or EuO/KTO) interfaces \cite{LAOEuOKTOSC2021, LAOKTO110SC2021} has reignited interest in this field. KTO-based interfaces feature interesting properties, including an intriguing dependence on the substrate orientation of the superconducting temperature ($\sim$2.2 K for KTO(111) and $\sim$0.9 K for KTO(110) interfaces) \cite{LAOEuOKTOSC2021, LAOKTO110SC2021}, in any case, higher than LAO/STO ($\sim$0.2 K) \cite{LAOSTOSC2007, LAOSTO110SC2014, LAOSTO111SC2017}. Analogous to the LAO/STO interface, where the Ti 3$d$ electrons are key to understanding the origin and the properties of the 2DEG~\cite{LAOSTOTi3d2008, LAOSTOTi3d2009}, Ta 5d electrons are supposed to play an important role in KTO-based interfaces \cite{KTOSurfTa5dARPES2012, KTOSurfTa5d2012, LAOKTOTa5d2017, LAOEuOKTOSC2021, LAOKTOTa5d2021}. In LAO/STO interfaces, the Ti $3d$ electrons are believed to be composed of localized and itinerant types. This is the key issue to understanding the origin of the interfacial 2DEG and their transport and magnetic properties \cite{LAOSTOcarrier2007, LAOSTOcarrier2008, LAOSTOcarrier2009, LAOSTOFMSC2011}. However, the nature of the Ta $5d$ electrons in KTO-based interfaces is still unknown, and uncovering this information will shed light on the 2DEG and superconductivity in these interfaces.

To reveal the nature of the interfacial electrons, resonant inelastic x-ray spectroscopy (RIXS) has proven to be a powerful tool in LAO/STO systems \cite{LAOKTORIXS2010, LAOSTORIXS2011, LAOSTORIXS2018, LAOSTORIXS2020} and holds great potential for analyzing KTO-based interfaces as well. In LAO/STO interfaces RIXS studies have revealed much higher carrier densities of 2DEG compared to the values obtained through Hall effect measurements, suggesting a coexistence of both localized and itinerant charge carriers \cite{LAOKTORIXS2010, LAOSTORIXS2018}. In bulk STO, Ti$^{4+}$ ions (with nominal $3d^0$ electronic configuration) do not produce $dd$ excitations, while a sizable contribution to the RIXS signal comes from Ti$^{3+}$ ions ($3d^1$) with localized electrons at the LAO/STO interface. Similarly, the valence of Ta is reduced at the LAO/KTO interface, as compared to the bulk KTO as evidenced by X-ray absorption spectroscopy (XAS) measurements \cite{LAOEuOKTOSC2021}; however, the degree of (de)localization is still a matter of debate. Here, we aim at addressing this issue and elucidating the nature of the 2DEG at the LAO/KTO interface.

Here we report our results of RIXS study on the KTO-based interfaces and electron-doped bulk KTO using Ta $L_3$ edge. The results showed a prominent charge transfer signal, but $dd$ excitations were not detected in any of the samples. By combining our RIXS results with XAS, transport measurements, and calculations based on a low-energy tight-binding model, we conclude that, unlike LAO/STO, localized electrons are either very few or totally absent in KTO-based interfaces.

\section{Experiments}

\begin{figure}[htbp]
\includegraphics[width=0.45\textwidth] {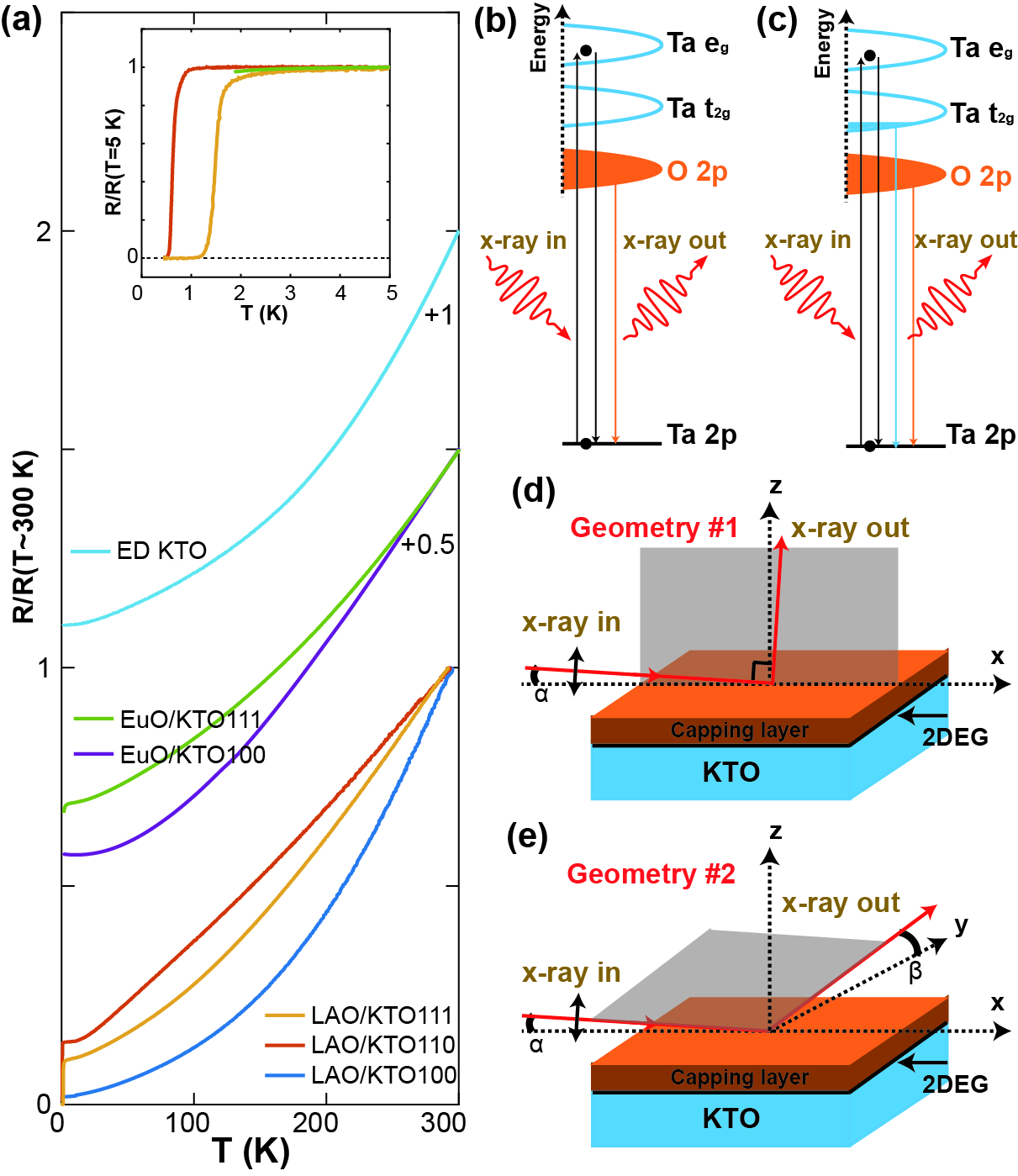}
\caption{(a) The resistance of samples versus temperature. The inset shows the resistance of superconducting samples below 5 K. (b) Schematic plot of the CT excitation in the insulating sample. (c) Schematic plot of the CT and $dd$ excitations in the electron-doped sample. (d) The geometry used in ESRF. (e) The geometry used in APS.}
\label{RTRIXS}
\end{figure}

The LAO/KTO interfaces with (100), (110) and (111) orientations were prepared by pulsed laser deposition. Laser ablation of a single crystalline LAO target was achieved by a 248-nm KrF excimer laser with fluence $\sim$1 J/cm$^2$ and a repetition rate of 2 Hz. The depositions were performed at 680 $^\circ$C, at mixed $1\times10^{-5}$ mbar O$_2$ and $1\times10^{-7}$ mbar H$_2$O vapor. The LAO film thickness was controlled at 4 nm for all samples. Then samples were cooled down to room temperature (RT) under the growth atmosphere. The EuO/KTO interfaces with (100) and (111) orientations were prepared via oxide molecular beam epitaxy at 500 $^\circ$C. A two-step process was used, which involved a $\sim$1 nm Eu buffer layer without oxygen, followed by a $\sim$ 3 nm EuO layer with an O$_2$ pressure of $\sim1\times10^{-9}$ mbar. After cooling to about RT, a 4 nm MgO layer was deposited $in situ$ by $e$-beam evaporation. The MgO layer can protect the EuO films when samples are transferred into the air.
The resistance and Hall effect of EuO/KTO samples were measured through a van der Pauw method. For LAO/KTO samples, the contacts to the interfaces were made by ultrasonic
bonding with Al wires and the four-probe DC method was used for all the resistance measurements. The KTO(111) and KTO(110) interfaces are superconducting, while KTO(100) samples are metallic, as illustrated in Fig. \ref{RTRIXS}(a). The itinerant carrier densities can be estimated by the Hall effect and are roughly $0.7-1.4\times10^{14}$/cm$^2$ for all interfaces (see Table \ref{Halleffect}).

The electron-doped bulk KTO sample (ED KTO) was prepared by a similar method depicted in Ref. \cite{KTOgrowth1965, KTOgrowth1967, KTOgrowth2009, KTOgrowthJJAP2009, KTOgrowth2010}. The K$_2$CO$_3$ (45 mmol), Ta$_2$O$_5$ (30 mmol) and BaCO$_3$ (0.5 mmol) powders were mixed in a platinum crucible, which was covered with a lid and placed in an Al$_2$O$_3$ crucible. The extra K$_2$CO$_3$ was introduced to maintain an excess of K$_2$O and serve as a flux. The BaCO$_3$ served as an electron dopant by substituting K$^{+}$ by Ba$^{2+}$. The mixture was heated to 950 $^\circ$C for 2 hours for pre-chemical reactions. Then the mixture was heated to 1400 $^\circ$C, held at this temperature for 2 hours, cooled to 900 $^\circ$C by a rate of 20 $^\circ$C/h and then to RT quickly. The cooling rate was fairly high and would introduce oxygen vacancy in the produced KTO samples due to the insufficient reactions with oxygen in the air \cite{KTOgrowth1965, KTOgrowth1967}. Thus, the chemical formula of produced KTO samples should be (K$_{1-x}$,Ba$_x$)TaO$_{3-y}$, with electron-doped by Ba$^{2+}$ and oxygen vacancy. The produced doped KTO samples were black and showed metallic behavior by measuring resistance versus temperature. From Hall measurement, the doped level was estimated to be about 0.001 electrons per Ta site (see Table \ref{Halleffect}).

\begin{table*}[htbp]
\caption{Summary of the samples studied in this work and their carrier density at room temperature. The results from STO samples \cite{LAOSTORIXS2020} are included for comparison.}
\begin{ruledtabular}
\begin{tabular}{lccccc}
Sample  & T$_c$  & n$_{3D}$ (cm$^{-3}$) & n$_{2D}$ (cm$^{-2}$) & thickness & electrons per Ta/Ti\\
\hline
LAO/STO \cite{LAOSTORIXS2020} & $\sim$0.2 K & $2-4\times10^{19}$ & $2-3\times10^{13}$  &  8 nm & 0.001-0.002  \\
LAO(EuO)/KTO(111) & $\sim$2.2 K & $1-2\times10^{20}$ & $7-14\times10^{13}$ & 7 nm & 0.006-0.012 \\
ED STO \cite{LAOSTORIXS2020} & & $5\times10^{19}$ & & & 0.003 \\
Insulating STO \cite{LAOSTORIXS2020} & & $<10^{15}$ & & & $<10^{-7}$ \\
ED KTO & & $2\times10^{19}$ & & & 0.001 \\
\end{tabular}
\end{ruledtabular}
\label{Halleffect}
\end{table*}

The RIXS and XAS measurements were performed at the ID20 beamline of the European Synchrotron Radiation Facility (ESRF) \cite{MorettiSala:il5001} and the 27-ID-B beamline of the Advanced Photon Source (APS). The energy resolutions (full width at half maximum) were 0.3 eV at ESRF while 0.7 eV or 0.12 eV at APS. The KTO interfaces and bulk sample were investigated at Ta $L_3$ edge. XAS was obtained at RT, while RIXS measurements were carried out at both RT and 20 K. The experimental geometries are shown in Fig. \ref{RTRIXS}(d) (for ESRF) and \ref{RTRIXS}(e) (for APS). Grazing incident angle was used in both geometries to minimize the x-ray penetration depth and enhance the interfacial signals. The estimated detection depth was a few nanometers when the incident angle ($\alpha$) was less than 0.2$^\circ$ and $\sim$100 nm when $\alpha=1^\circ$. In the ESRF (APS) geometry, the scattering plane was perpendicular (nearly parallel) to the sample surface and indicated by a shadow plane. The incident photon polarization was within the xz plane, which was horizontal in our experiments.

The XAS spectra were obtained at the Ta $L_3$ edge, where an electron would be excited from the Ta $2p$ shell to the Ta $5d$ orbital. Two types of XAS were performed in our experiment: total fluorescence yield (TFY) and partial fluorescence yield (PFY). The PFY is a core-to-core RIXS process. After the excitation of an electron from the $2p_{3/2}$to $5d$ orbital, a decay via the $3d_{5/2}$ to $2p_{3/2}$ transition ($L_{\alpha1}$) followed coherently. The PFY mode usually produces well-separated peaks for $5d$ transition metal systems due to a smaller number of Auger decay possibilities \cite{PFY2017}.

RIXS is a photon-in, photon-out process, where an electron is excited from the Ta $2p$ shell to the Ta $5d$ orbital at Ta $L_3$ edge. After the excitation, different radiative decay channels exist. The scattering process can be elastic, with the final state being the same as the original ground state. Alternatively, the emitted photon can have a lower energy (inelastic scattering) and an excited final state. The excited final state could involve one electron in the Ta $5d$ orbital and one hole in the O $2p$ orbital, which is referred to as charge transfer (CT) type excitation. This scenario is depicted in an insulating KTO (see Fig.~\ref{RTRIXS}(b)). When there are occupied electrons in Ta $5d$ orbital in the initial state, an alternative excited state would involve an electron in an unoccupied Ta $5d$ orbital and one hole in occupied Ta $5d$ states, referred to as $dd$ excitation. The CT and $dd$ excitations are supposed to be observed in an electron-doped KTO (see Fig.~\ref{RTRIXS}(c)).

\begin{figure}[htbp]
\includegraphics[width=0.45\textwidth] {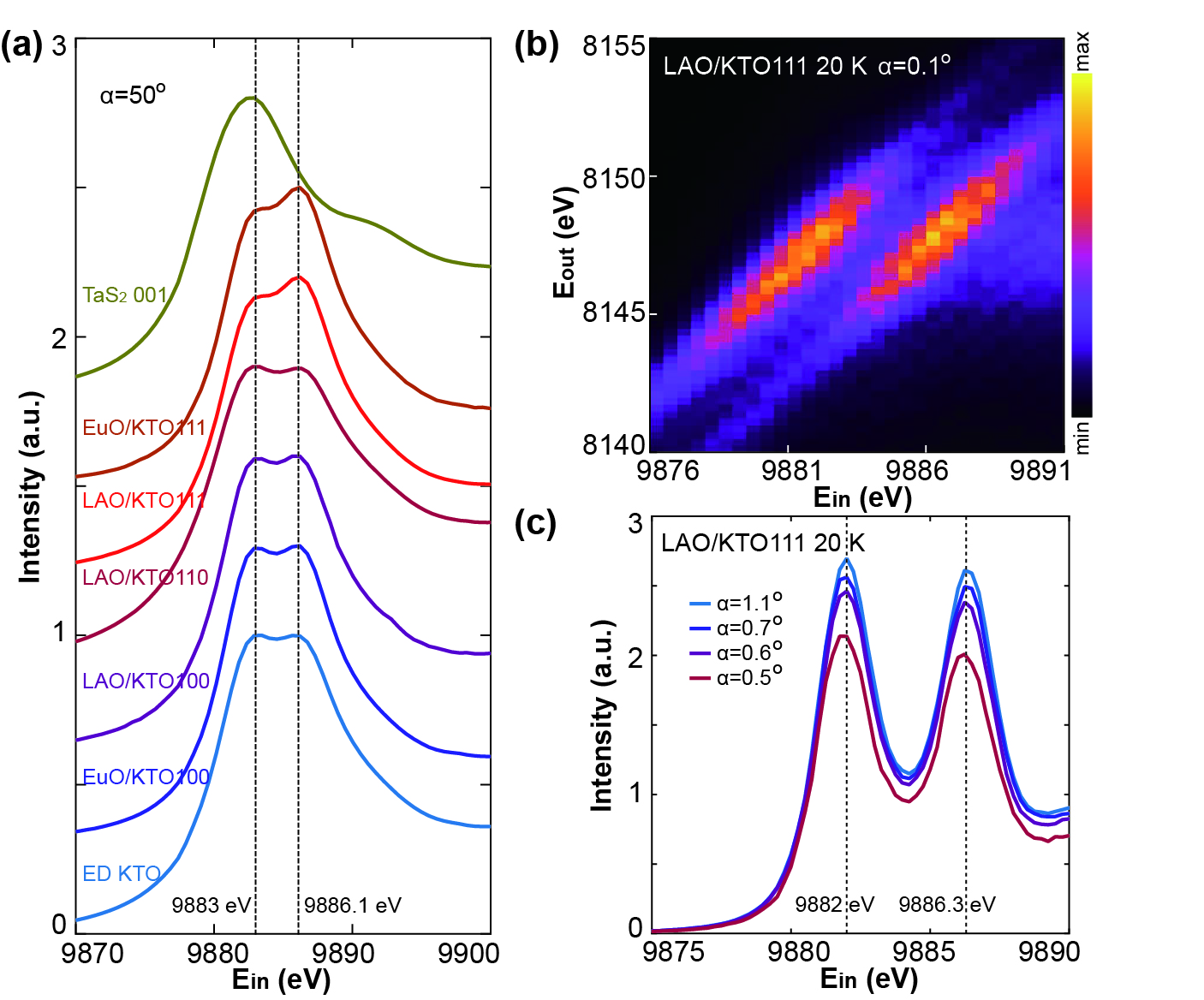}
\caption{(a) The TFY XAS of all samples measured in ESRF. (b)(c) The PFY XAS of LAO/KTO(111) measured at APS.}
\label{XAS}
\end{figure}

\section{Results}
\subsection{XAS}

The TFY and PFY of all samples are shown in Fig. ~\ref{XAS}. For TFY XAS, the data were obtained at ESRF with an incident angle of $\alpha=50^o$ and at RT. The large incident angle ensures that bulk information is recorded. The KTO samples, regardless of whether they are interfaces or electron-doped bulk, show similar double-peak features. These double peaks are located at 9883.0 eV and 9886.1 eV, respectively, and are characteristic of Ta$^{5+}$ oxide state \cite{LAOEuOKTOSC2021}. In contrast, the XAS of the TaS$_2$ sample (see Fig.~\ref{XAS}(a)) shows a single peak with a peak position slightly lower than 9883.0 eV. This single peak feature in XAS has also been observed in another Ta compound with a $5d^1$ electron configuration \cite{Cs2TaCl6XASRIXS2019}. In comparison, the XAS of ED KTO is not similar to TaS$_2$ but is instead similar to the KTO interfaces where the bulk is an insulating KTO. This suggests the lack of $5d^1$ electrons in ED KTO.

Due to the broad peaks in the TFY XAS mode, it is difficult to accurately determine the peak positions and potential small changes in the XAS. To better resolve the peaks, we tried the PFY XAS mode at APS, which produces well-separated peaks for $5d$ transition metal systems \cite{PFY2017}. Here we focus on the LAO/KTO(111) interface at 20 K. With an incident angle of approximately 0.1$^o$, the map of PFY XAS intensity was obtained as a function of the incident and outgoing photon energy (as seen in Fig.~\ref{XAS}(b)), which displays clearly two separated peaks. By focusing on a constantly emitted photon energy window, we can see a sharper and better-resolved feature compared to the TFY mode, as shown in Fig.~\ref{XAS}(c). Two peaks are located at 9882.0 eV and 9886.3 eV with an interval of 4.3 eV, which is attributed to the crystal field splitting between $t_{2g}$ (left peak) and $e_g$ (right peak) manifolds. By decreasing the incident angle, the energy peak positions of XAS show a weak redshift, indicating a slight reduction in Ta valence at the interface compared to the bulk. A similar result was previously reported in the TFY XAS of KTO interfaces \cite{LAOEuOKTOSC2021}.

\subsection{RIXS of KTO interfaces}
\begin{figure}[htbp]
\includegraphics[width=0.45\textwidth] {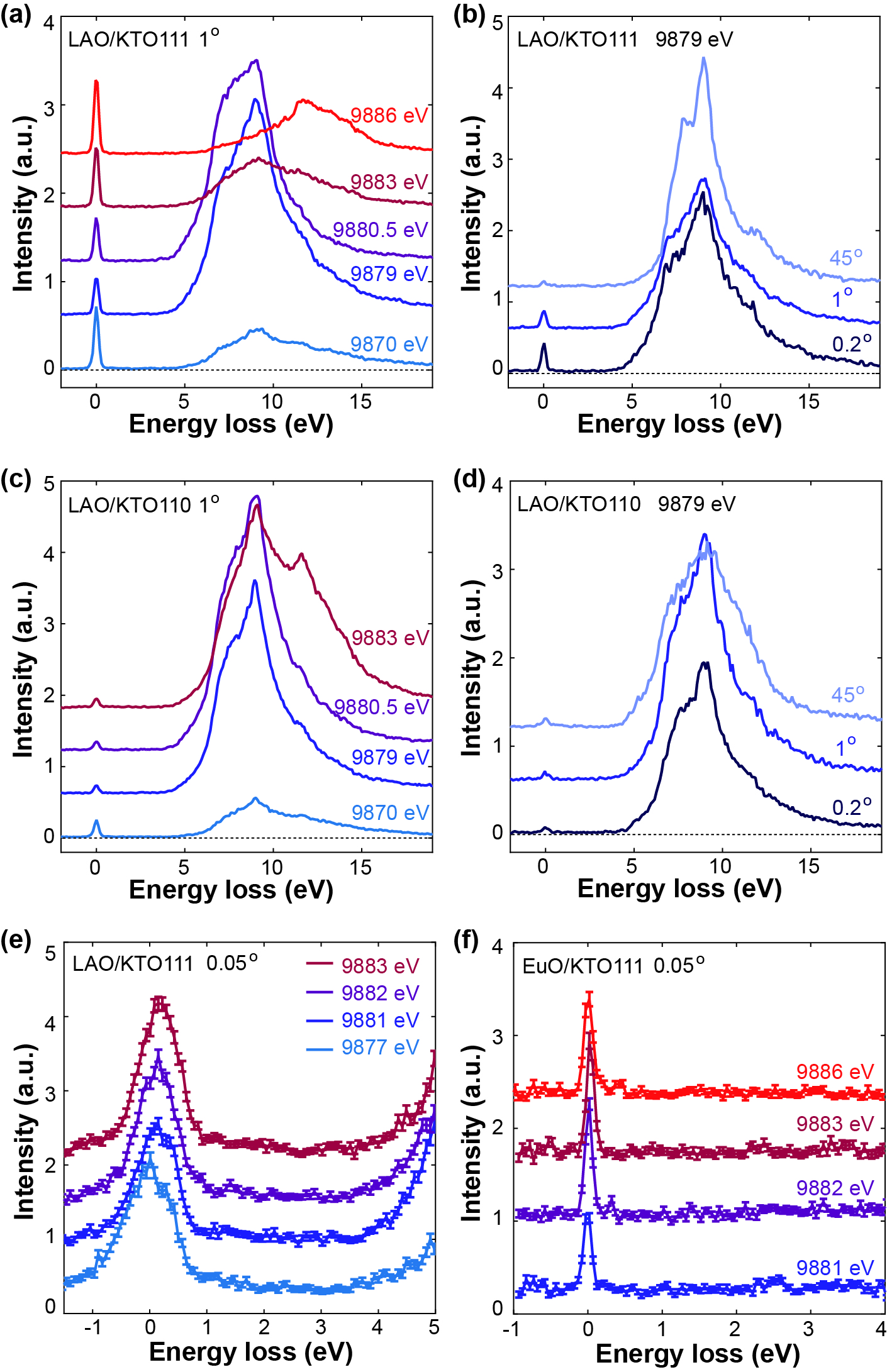}
\caption{The RIXS spectra of (a,b,e) LAO/KTO(111), (c,d) LAO/KTO(110) and (f) EuO/KTO(111) interfaces. Parameters for the spectra: (a,c) $\alpha=1^o$, energy resolution 0.3 eV, varied incident energy, geometry 1, (b,d) $E_{in}=9879$ eV, energy resolution 0.3 eV, varied $\alpha$, geometry 1, (e) $\alpha=0.05^o$, energy resolution 0.7 eV, varied incident energy, geometry 2, (f) $\alpha=0.05^o$, energy resolution 0.12 eV, varied incident energy, geometry 2. Data are vertically offset for clarity.}
\label{RIXSinterface}
\end{figure}

RIXS was employed to investigate the electronic properties, specifically the localization or delocalization of electrons, in the superconducting KTO interfaces. We first focus on the LAO/KTO(111) interface. The RIXS spectra were collected at RT with the geometry shown in Fig. \ref{RTRIXS}(d). As illustrated in Fig.~\ref{RIXSinterface}(a), by increasing the incident photon energy from the $t_{2g}$ to the $e_g$ peak in the XAS profile, we observe a clear CT signal. The CT signal starts at around the energy loss $\sim$ 4.0 eV, close to the band gap of KTO as determined by optical methods \cite{KTOGap1965,KTOGap1967}. Two distinct charge transfer (CT) excitations were observed, resonant at different incident x-ray energies. Specifically, when the incident x-ray was resonant at the Ta $t_{2g}$ (or $e_g$) peak, the CT excitation was primarily located around 8 eV (or 12 eV), indicating the transfer of electrons from the O $2p$ band to the Ta $t_{2g}$ (or $e_g$) band. However, irrespective of the incident photon energy, we did not observe any $dd$ excitations below 4 eV energy loss. In LAO/STO, $dd$ excitations have been observed at $\sim$ 2.5 eV \cite{LAOKTORIXS2010, LAOSTORIXS2011, LAOSTORIXS2018, LAOSTORIXS2020}. To rule out the influence of the x-ray penetration depth, we changed the incident angle while fixing the incident energy at 9879 eV. Still, only CT signals were present (see Fig. ~\ref{RIXSinterface}(b)). The lineshape of the CT signal with $\alpha=45^o$ is different from those with the grazing-in angle, likely due to momentum-dependent RIXS spectra. We then studied the superconducting LAO/KTO(110) interface and obtained similar results, as shown in Fig.~\ref{RIXSinterface}(b) and (d). We observed clear CT signals but no $dd$ excitations. Again, the different lineshape from the LAO/KTO(111) interface is likely due to the different momenta.

To rule out the possibility of low incident x-ray flux, we conducted additional measurements on the LAO/KTO(111) interface using higher flux and lower energy resolution (0.7 eV). We collected RIXS spectra with energy losses of less than 5 eV, where $dd$ excitations were expected, under the geometry shown in Fig. \ref{RTRIXS}(e), but we did not observe any $dd$ excitations. Further improvement of the energy resolution to 0.12 eV did not help either. As illustrated in Fig. \ref{RTRIXS}(f)), RIXS spectra of EuO/KTO(111) also showed no evidence of $dd$ excitations with an energy resolution of 0.12 eV.

In conclusion, the $dd$ excitations were consistently absent in the RIXS spectra, regardless of the orientation of the interface, the type of overlayer, the measurement geometry, or the energy resolution. The results suggest a very low or complete lack of localized Ta $5d$ electrons at the KTO interface.

\subsection{RIXS of electron-doped KTO bulk samples}

\begin{figure}[htbp]
\includegraphics[width=0.35\textwidth] {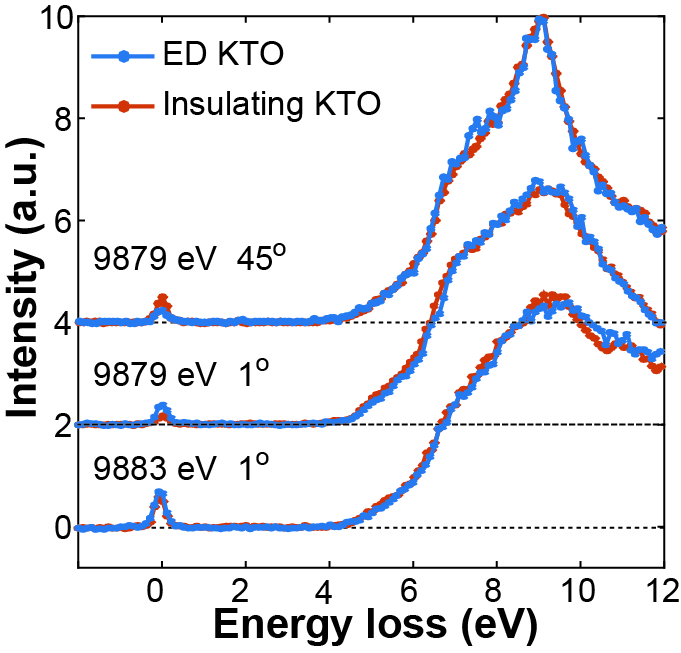}
\caption{The RIXS spectra of ED KTO (blue curves) and insulating KTO (orange curves), collected at geometry 1. The spectra are scaled and shifted vertically for comparison.}
\label{RIXSbulk}
\end{figure}

The KTO interfaces are electron-doped, so we also conducted RIXS measurements on the electron-doped KTO bulk materials for comparison. We varied the incident x-ray energies  ($E_{in}=$9879 eV, 9883 eV) and incident angles ($\alpha=1^o, 45^o$) during RIXS measurements, as shown in Fig.~\ref{RIXSbulk}. The $dd$ excitations were still not present in the ED KTO bulk samples. 
Furthermore, by comparing the RIXS spectra of the electron-doped KTO to that of the insulating KTO, the spectra of both samples can be scaled well together. This indicates that even in the ED KTO bulk materials, the density of Ta $5d$ electrons is too low to be detected.

\subsection{RIXS calculations}
\begin{figure}[htbp]
\includegraphics[width=0.5\textwidth] {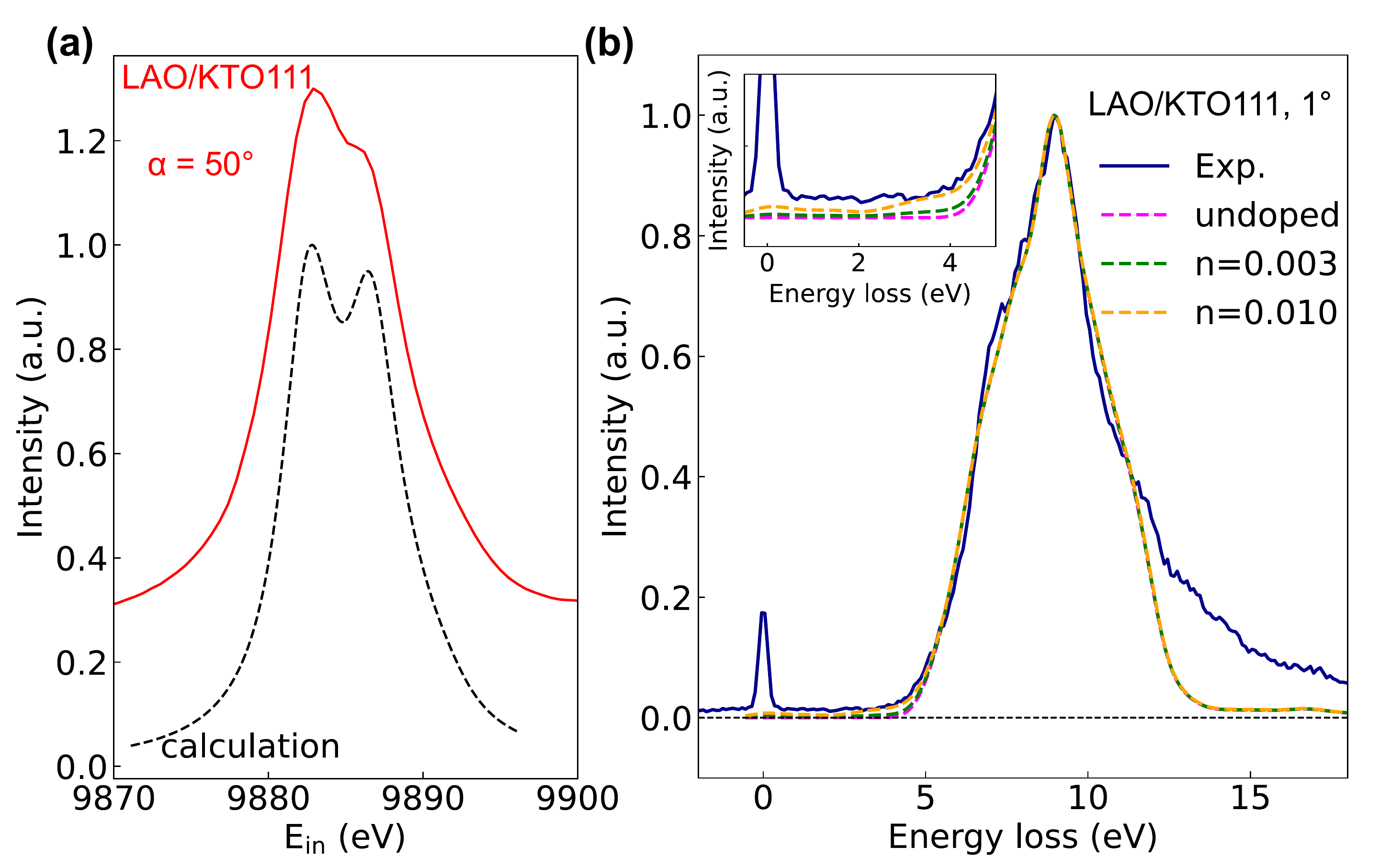}
\caption{(a) Comparison between experimental XAS and calculated XAS spectra. We have subtracted a tanh-function background for better comparison. (b) Calculated RIXS spectra with $E_{in}=9879$ eV for LAO/KTO111 with different electron densities per Ta. The corresponding experimental spectrum in Fig.~\ref{RIXSinterface}(a) is shown for comparison. The inset zooms into the low-energy region for better visibility.}
\label{fig:rixs_cal}
\end{figure}

To quantitatively understand the RIXS spectra of pristine and doped KTO, we calculated the RIXS cross section at the Ta $L_3$ edge using a low-energy tight-binding model $H_\mathrm{TB}$ assuming fully itinerant Ta-5$d$ and O-2$p$ electrons. The model is constructed using Wannier orbitals obtained from density-functional theory (DFT) Kohn-Sham orbitals~\cite{wien2k_a,wien2k_b,wannier90}. Such a model is capable of capturing $dd$ and charge transfer excitations on an equal footing. Note that an additional shift of the Ta-5$d$ orbitals is included in $H_\mathrm{TB}$ to compensate for the underestimated band gap in DFT. The RIXS cross-section is then given as
\begin{equation}
    I_\mathrm{RIXS}(\vb{q}, \omega,\omega_\mathrm{in}) = -\frac{1}{\pi} \Im \ev*{R^\dag_{\vb{q}} \frac{1}{\omega - H_\mathrm{TB} + \mathrm{i} 0^+} R_{\vb{q}} },
\end{equation}
where the RIXS operator reads
\begin{equation*}
    R_{\vb{q}} = \sum_{j\alpha\beta} e^{\mathrm{i} \vb{q}\cdot \vb{r}_j} \chi_{\alpha\beta} d_{j\beta} \frac{1}{\omega_\mathrm{in}-H_c+\mathrm{i} \Gamma} d^\dag_{j\alpha}.
\end{equation*}
The fermionic operator $d^{(\dag)}_{j\alpha}$ annihilates (creates) an electron in the $d$ orbital labeled by its spin-orbital index $\alpha$ at site $j$. The dipole matrix elements $\chi_{\alpha\beta}$ depend on the experimental geometry and light polarization. While the valence-valence interactions are omitted in our calculation considering the weakly interacting nature of $5d$ electrons, a core-valence interaction of 2.0 eV is included in the intermediate state Hamiltonian $H_c$ to reproduce the experimental XAS lineshape, as shown in ~\ref{fig:rixs_cal}(a). In addition, the inverse core-hole lifetime $\Gamma$ is set to 2.5 eV to match the observed broadening of XAS. Due to this large broadening, the spectral properties converge quickly with the number of $\vb{k}$ points. In our calculation, the Brillouin zone is sampled on a $\vb{k}$-grid of size $10\times 10\times 10$.

Figure~\ref{fig:rixs_cal} (b) shows the calculated RIXS spectra with different doping concentrations. Spectra of different doping levels are obtained by mixing the undoped spectrum with a spectrum calculated with two electrons per Ta. For electron density below $0.010$ per Ta, the calculated spectra show a weak intensity in the gaped region ($\lesssim$ 5 eV) below the experimental background, in accord with the estimated carrier density in Tab.~\ref{Halleffect}. The good agreement between the DFT calculations based on a pure itinerant model and our RIXS spectra confirms that the KTO samples lack local $5d^1$ electrons.

\begin{figure}[htbp]
\includegraphics[width=0.35\textwidth] {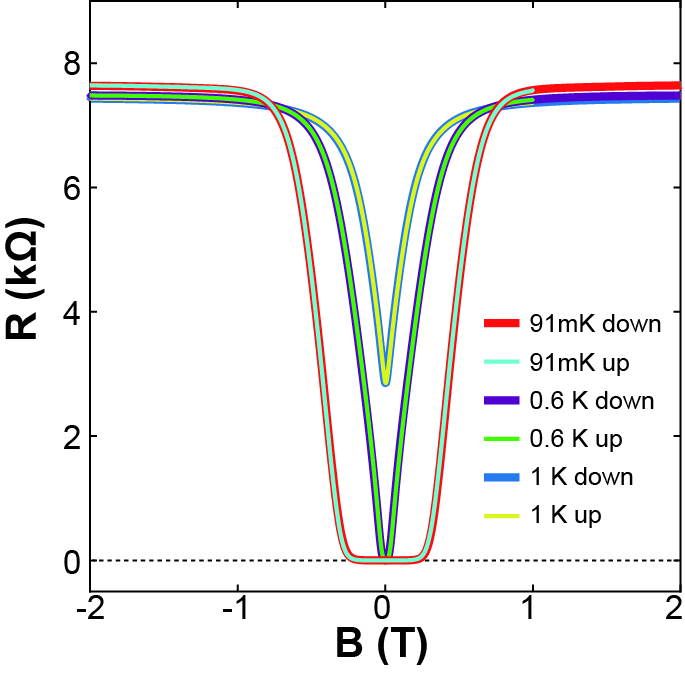}
\caption{The resistance of EuO/KTO(111) versus magnetic field.}
\label{MRKTO}
\end{figure}

\subsection{Magneto-resistance measurement}
The lack of Ta $5d^1$ electrons can help explain why there is no magnetic hysteresis in our magneto-resistance (MR) measurements. As shown in Fig.~\ref{MRKTO}, when the magnetic field perpendicular to the interface is scanned downwards first and then upwards, the resistance repeats itself regardless of the measuring temperature. In contrast, MR in the LAO/STO interfaces exhibits hysteresis \cite{LAOSTOSCFM2011, LAOKTO110MR2016, LAOKTO111MR2017}. Hysteresis typically indicates the presence of magnetic order, which requires local moments. In STO-based interfaces, the local moments are provided by local Ti $3d$ electrons. However, in KTO-based interfaces, due to the absence of local Ta $5d$ electrons, local moments do not exist, leading to the lack of hysteresis in MR measurements.

\section{Discussion}
To deepen our understanding of electron behavior in both KTO-based interfaces and KTO bulk materials, we compared our findings with previous studies conducted on STO. Our results indicate that the KTO interfaces have a higher density of itinerant electrons than the STO interfaces, as measured by the Hall effect  (see Table \ref{Halleffect}). However, while the RIXS spectra of STO interfaces exhibit  $dd$ excitations, which are attributed to the high density of localized Ti $3d^1$ electrons \cite{LAOKTORIXS2010, LAOSTORIXS2011, LAOSTORIXS2018, LAOSTORIXS2020}, the absence of such excitations in KTO interfaces suggests that localized Ta $5d^1$ electrons are either absent or present in very low density.
To confirm this observation, we performed simulations and found that the density of itinerant Ta $5d$ electrons is close to the detection limit for $dd$ excitations. If localized Ta $5d^1$ electrons were present, they would contribute to  $dd$ excitations and make them more visible. Since we did not observe any $dd$ excitations in our experiments, we can conclude that localized Ta $5d^1$ electrons are missing in KTO interfaces.

The absence of localized Ta $5d^1$ electrons in KTO-based interfaces may contribute to the unique properties of KTO. In electron-doped KTO, which has a slightly lower electron density than electron-doped STO, $dd$ excitations again disappear, indicating that localized Ta $5d^1$ electrons do not form. The larger spatial extension and stronger delocalization of Ta $5d$ electrons compared to Ti $3d$ electrons may contribute to this absence of localized electrons.

Considering that the $T_c$ (superconducting critical temperature) is significantly higher in LAO/KTO than LAO/STO, the presence of itinerant electrons in KTO may play a role in the higher $T_c$. Localized $d$ electrons can give rise to local magnetic moments that suppress superconductivity, so the absence of such electrons in KTO could contribute to the enhanced superconductivity observed in this material system.

\section{Summary}
We have conducted a comprehensive investigation into KTO-based systems with varying orientations, capping layers, and electron doping. Through the use of resonant inelastic x-ray spectroscopy (RIXS), we found no evidence of $dd$ excitations in samples with up to $\sim 0.01e$ electrons per Ta site. Our calculations, which assumed purely itinerant Ta $5d$ electrons, agree well with the RIXS results. Although further experiments with higher carrier densities are needed to fully solidify this observation, the consistency between the data and calculations points towards the absence of localized Ta $5d^1$ electrons. This finding agrees with the lack of magnetic hysteresis in magneto-resistance measurements, advancing our understanding of these fascinating materials.

\section{ACKNOWLEDGMENTS}

Y.Y.P. is grateful for financial support from the Ministry of Science and Technology of China (2019YFA0308401 and 2021YFA1401903) and the National Natural Science Foundation of China (11974029). Y.L. acknowledges support from the National Key R$\&$D Program of China under grant No. 2022YFA1403000 and the National Natural Science Foundation of China under grant No.12274207. J.N. is grateful for financial support from the Strategic Priority Research Program of Chinese Academy of Sciences (Grant No. XDB28000000) and the National Natural Science Foundation of China (11921005). We acknowledge the European Synchrotron Radiation Facility (ESRF, France) for providing beamtime. 
This research used resources of the Advanced Photon Source, a U.S. Department of Energy (DOE) Office of Science User Facility operated for the DOE Office of Science by Argonne National Laboratory under Contract No. DE-AC02-06CH11357.

\end{document}